\documentclass[pra,aps,twocolumn,superscriptaddress,notitlepage]{revtex4-1}
\usepackage{amssymb,amsfonts,amsmath,amstext,graphicx,hyperref}
\usepackage{xr-hyper}

\usepackage{tikz}
\usepackage{braket}
\usepackage{graphicx}
\newtheorem{theorem}{Theorem}

\usepackage[normalem]{ulem}
\usepackage{tabularx}
\def\bra#1{\langle{#1}|}
\def\ket#1{|{#1}\rangle}
\def\braket#1{\langle{#1}\rangle}

\usepackage{braket}
\usepackage{mathtools}
\hypersetup{
	colorlinks=true,
	linkcolor=blue,
	filecolor=magenta,
        citecolor=blue,
	urlcolor=brown,
}
\usepackage{xcolor}

\begin{document}

\title{Projected Optimal Sensors from Operator Orbits}

\author{Sooryansh Asthana}
\affiliation{Department of Physics, Indian Institute of Technology Bombay, Powai, Mumbai 400076, India}
\author{Yeshma Ibrahim}
\affiliation{Department of Physics, Indian Institute of Technology Bombay, Powai, Mumbai 400076, India}
\author{Norman Tze Wei Koo}
\affiliation{Pritzker School of Molecular Engineering, The University of Chicago, Chicago, Illinois 60637, USA}
\author{Sai Vinjanampathy}
\email{sai@phy.iitb.ac.in}
\affiliation{Department of Physics, Indian Institute of Technology Bombay, Powai, Mumbai 400076, India}
\affiliation{Centre of Excellence in Quantum Information, Computation, Science and Technology, Indian Institute of Technology Bombay, Powai, Mumbai 400076, India}
\date{\today}

\begin{abstract}
We unify Ramsey, twist-untwist, and random quantum sensors using operator algebra and account for the Fisher scaling of various sensor designs. 
We illustrate how the operator orbits associated with state preparation inform the scaling of the sensitivity with the number of subsystems. 
Using our unified model, we design a novel set of sensors in which a projected ensemble of quantum states exhibits beyond-shot-noise metrological performance. We also show favorable scaling of Fisher information with decoherence models and loss of particles. 
\end{abstract} 
\maketitle
	
\textit{Introduction.---} 
Precision measurements are at the heart of foundational questions such as dark matter detection \cite{darkmatter,bass2024quantum}, electron electric dipole moment \cite{electronedm,roussy2023improved,raymolecules}, and exoplanet detection \cite{zixin}. 
Precision sensing also has tremendous applications in areas such as biology \cite{biomed,yukawa2025quantum}. In this context, quantum metrology seeks to surpass the sensitivity of classical sensors in estimating unknown parameters \cite{giovannetti2011advances,RevModPhys.89.035002,szczykulska2016multi,pezze2018quantum,sidhu2020geometric}. Local estimation, where unknown parameter shifts are detected about a known point, differs from global estimation \cite{mukhopadhyay2025saturable,boeyens2025role,mukhopadhyay2025current} strategies. For local quantum sensing with an unbiased estimator, the variance of the estimator defines the sensitivity and is lower bounded by the so-called quantum Cramér-Rao bound. For pure initial probe states, this variance is bounded from below by the inverse of the quantum Fisher information (QFI), which in turn is proportional to the variance of the generator in the initial state. Hence, one of the design paradigms in sensing has been the preparation of suitable probe states from a fiducial state. 

Sensor design has evolved by optimizing the initial state preparation \cite{PhysRevA.54.R4649, PhysRevA.55.2598, PhysRevA.66.023819}, the phase-embedding Hamiltonian \cite{napolitano2011interaction, vetrivelan2022near, PhysRevA.111.L020402}, and measurements \cite{PhysRevLett.72.3439, pezze2014quantumtheoryphaseestimation,nair,zixin}. State optimization focused on generating \textit{fine-tuned} quantum states such as path entangled states or cat states \cite{giovannetti2004quantum,resch2007time, afek2010high, ono2013entanglement,   cao2024multi, kim2025distributed}. Owing to their sensitivity to decoherence \cite{huelga1997improvement}, theoretical and experimental focus shifted to beyond shot-noise sensors, which are also experimentally feasible. This includes the so-called twist-untwist protocol \cite{davis_approaching_2016} and analogues \cite{PRXQuantum.6.010304}, which have achieved considerable experimental success \cite{PhysRevResearch.7.013227, PhysRevResearch.6.023179, PhysRevA.111.042611, franke2023quantum}. Motivated by independent developments exploring the role of dynamics to entanglement generation, kicked-top and other Hamiltonians exhibiting quantum chaos emerged as candidates \cite{fiderer2018quantum, doi:10.1126/science.adg9500, shi2024quantum, zou2025enhancing} displaying beyond-shot noise Fisher scaling at intermediate times \cite{cotler2017chaos}. 
We refer to such designs in which state preparation is not fine-tuned but instead exploits the statistical properties of time-evolved states as \textit{resilient} designs.  
A common theme in resilient sensor design is the presence of permutational invariance, as noted in \cite{oszmaniec_random_2016} where authors proposed scrambling as the resource for enhanced sensing. Recent proposals \cite{kobrin2024universal} of random sensor designs exclude such symmetries, highlighting the need for a unifying theory that goes beyond symmetries.

\begin{figure}
    \centering
    \includegraphics[width=0.90\columnwidth]{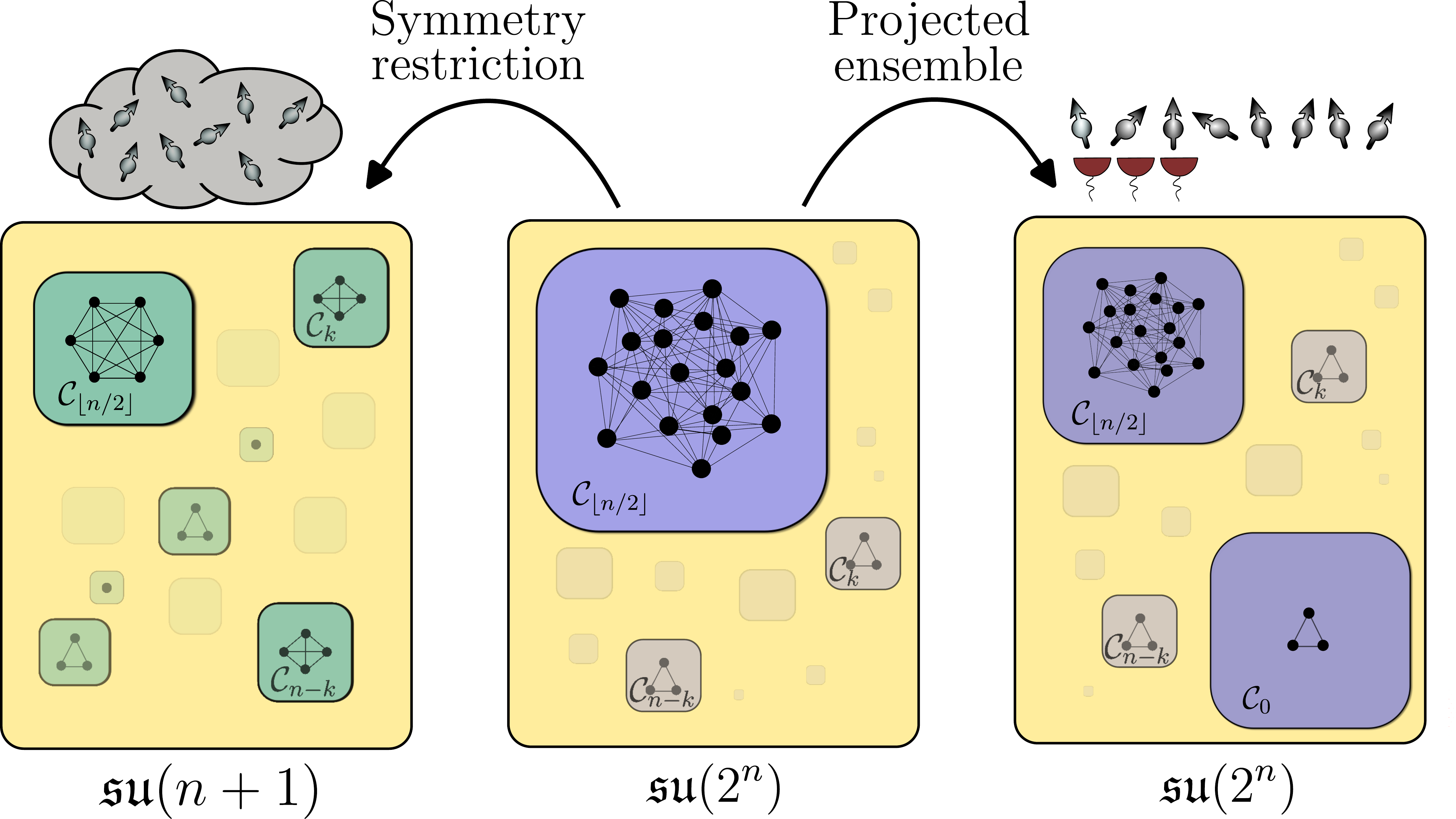}
    \caption{An edge connects two Pauli strings that are equivalent to each other in the way they anticommute with the generator of phase embedding, thereby defining an equivalence class $\mathcal{C}_k$. Dynamical constraints, either via symmetry (left) or measurement and postselection (right), prevent featureless scrambling within the Hilbert space. This restricts the dynamics to metrologically sensitive operator orbits that have support over macroscopically distant equivalence classes.}
    \label{fig:dla}
\end{figure}

Hence, an inadequacy of the current theoretical efforts is a unifying theory that highlights the role of symmetries of the 
state preparation unitary in the subsequent metrological performance. 
In this manuscript, we unify all existing metrology schemes, including fine-tuned state preparation, twist-untwist protocols, and random sensing protocols with and without symmetries. We do so by showing that the \textit{Dynamical Lie Algebra} (DLA), or more generically the \textit{operator orbit} that generates resource states together with the properties of the phase embedding Hamiltonian is key to understanding the improved performance of all aforementioned sensors. We define the notion of \textit{metrologically compatible subspaces} which can be used to engineer these operator orbits.  
We show that by imposing broad restrictions, such as symmetries, on the class of dynamics, it is possible to constrain the DLA to subspaces that exhibit high average metrological performance. One of our key findings is that this is not the only way to constrain operator orbits to yield a large Fisher advantage. We demonstrate this by designing novel sensors whose resilience stems from postselection, rather than symmetries.
 
An $n$-qubit unitary that generates the initial state from a fiducial state $\ket{\psi_0}$, namely $\ket{\psi_{in}}=U\ket{\psi_0}$, can be written in terms of a \textit{composing Hamiltonian} $H_c$, with $U=\exp({-iH_c t})$. 
We take the phase-embedding generator to be a sum of $1$-local operators, $G = \sum_{i = 1}^{n}G_i$, and fix the fiducial state to be $\ket{0}^{\otimes n}$ \cite{Maccone06}. 
For dynamics generated by integrable composing Hamiltonians, it can be shown that generating resourceful states that saturate the Fisher information requires fine-tuning in their operator orbit (see Appendix~\ref{app: integrable}). 
In the remainder of this manuscript, we consider protocols using generic nonintegrable systems. 
This allows us to sample unitaries via the Haar measure on the Lie group generated by the composing Hamiltonians in the intermediate time regime \cite{cotler2017chaos}. 
Here, the ensemble-averaged Fisher information then serves as a measure of the metrological performance under a given set of controls in the composing Hamiltonian. 
It was previously noted that sampling from $\text{SU}(2^n)$ does not produce quantum-enhanced metrological advantage \cite{oszmaniec_random_2016}, an outcome attributed to the scrambling of quantum information \cite{kobrin2024universal}, a phenomenon we now qualify using DLAs \cite{larocca2022diagnosing, ragone2024lie, wiersema2024classification, dla-goh, graph_kavan}. 
Symmetry imposition on the class of dynamics constrain the DLA to subspaces with high average metrological sensitivity, as illustrated in Fig.~\ref{fig:dla}. 
Furthermore, we demonstrate that one can recover metrological performance from otherwise insensitive ensembles by constraining operator orbits using local projective measurements and postselection. 
These protocols enable sensor designs that surpass the standard quantum limit without requiring fine-tuned operator orbits.

\noindent\textit{Role of DLA in Metrology.---} 
Consider the $n$-qubit composing Hamiltonian, $H_c = \sum_{h_i \in \mathcal{G}} c_i (t) h_i$ with time-dependent coefficients $c_i$, constructed from the Hermitian dynamical generating set $\mathcal{G} = \{ h_1, h_2, h_3, ...\}$, where $|\mathcal{G}|\geq 2$. 
We consider scenarios in which $\mathcal{G} \subseteq \mathcal{P}_n = \{I, X, Y, Z\}^{\otimes n}$ is a set of Pauli strings, that define the class of dynamics of our interest. 
The DLA associated with this set is the closure of $i\mathcal{G}$ under commutation and real linear combinations, which forms a vector space over $\mathbb{R}$,
\begin{align}
    \mathfrak{g} &= \langle i\mathcal{G}\rangle_{\text{Lie}} \nonumber\\
 &= \text{span}_{\mathbb{R}} \bigl\{i\mathcal{G} \cup \big [ i\mathcal{G}, i\mathcal{G}\big] \cup  \big[i\mathcal{G}, \big [ i\mathcal{G}, i\mathcal{G}\big]\big] \cup ... \bigl\}.
\end{align}
The set of all unitaries that this algebra can generate is a connected Lie group, $\mathcal{U} = \text{exp} (\mathfrak{g})$. 
Each unitary in this group can be expressed as a sum of Pauli strings. 
The entire Lie group is accessible when the coefficients $c_i$, corresponding to the control Hamiltonians $h_i$, are independently controllable.  
For a given composing Hamiltonian, the unitary orbit generates all reachable states under its evolution that act as potentially resourceful probe states for metrology. 
We restrict ourselves to the case of time-independent composing Hamiltonians; however, our discussion applies in principle to the more general time-dependent case.
To construct metrologically resourceful probe states for a family of composing Hamiltonians, we develop below a notion of metrologically sensitive operator orbits. 

For any Pauli string $P = \bigotimes_{i = 1}^{n} P_i \in \mathcal{P}_n$, denote $N(P, G)$ as the number of sites $i$ at which $G_i$ anticommutes with the corresponding local action of $P$. 
We introduce an equivalence relation on $\mathcal{P}_n$ by imposing that $P \sim_G Q$ if and only if $N(P, G) = N(Q, G)$. Under such an equivalence relation, $\mathcal{P}_n$ can be divided into $n+1$ equivalence classes such that for each $k = 0, 1, 2, ...n$, the corresponding equivalence class is
\begin{equation} \label{eq:eq_class}
    \mathcal{C}_k(G) =\bigl\{P: N(P, G) = k\bigl\} .
\end{equation}
Using this, we define a \textit{metrologically compatible subspace} with respect to a reference Pauli string $P \in \mathcal{C}_p$ as
\begin{equation}
    \mathfrak{M}_p =  \text{span}_{\mathbb{C}}\Bigg(\bigcup_{|k - p| \sim \mathcal{O}(n)} \mathcal{C}_{k}\Bigg).
\end{equation}
This subspace contains linear combinations of all Pauli strings whose commutation structure with $G$ differs from $P$ on a macroscopic number of sites. 
Any linear combination in which $P$ appears with finite amplitude together with elements of $\mathfrak{M}_p$ generates a \textit{metrologically sensitive operator orbit} exhibiting quantum-enhanced sensitivity to the phase embedding generated by $G$. To illustrate this, consider a composing Hamiltonian $H_c = X^{\otimes n}$. 
This creates an operator orbit $\cos(t)I + i\sin(t)X^{\otimes n}$ which is a linear combination of $I \in \mathcal{C}_0$ and $X^{\otimes n} \in \mathcal{C}_{n}$, thereby producing the paradigmatic GHZ-like state, known for its metrological sensitivity, for almost all $t$.
In the next section, we use this notion to understand Haar random protocols in different subspaces.

\noindent \textit{Haar-Ramsey protocol.---}
We consider a local metrology protocol for an $n$-qubit probe, where an unknown real parameter $\theta \in \mathbb{R}$ is embedded via a unitary of the form $U_{\theta} = e^{-i\theta G}$. Composing Hamiltonians within the subalgebra ${\mathfrak{su}(d)}$ generates an initial state for the protocol at time $t$, which is then embedded with the parameter. 
This is followed by a time-reversed evolution under the composing Hamiltonian such that $\ket{\psi_\theta} = \overline{U}e^{-i\theta G} U \ket{\psi_0}$, where $U = e^{-iH_ct}$. Due to its similarity to the Ramsey protocol, but now incorporating Haar rotations, we refer to this scheme as the \textit{Haar-Ramsey} protocol.
A measurement onto the initial probe state and its orthogonal complement $\{\ket{\psi_0}\bra{\psi_0}, I-\ket{\psi_0}\bra{\psi_0}\}$ then saturates the QFI for pure states \cite{pezze2014quantumtheoryphaseestimation}.

\noindent\textit{Theorem 1.---(Average QFI and Lie orbit)}  
Let $\mathfrak{g}  = \oplus_{i} \mathfrak{g}_i \subseteq \mathfrak{su}(2^n)$ be a reductive subalgebra. The $n$-qubit Hilbert space then admits a decomposition $\mathcal{H} = \oplus_{i} \mathcal{H}_i$, where $ \mathcal{H}_i$ are the invariant subspaces of dimensions $d_i$ under the action of $\mathfrak{g}$, characterized by projections $\Pi_i$. For a fiducial probe state $\ket{\psi_0} \in \mathcal{H}_i$ and a given generator of the unitary phase encoding, $G$, the Haar-average QFI over the unitary group $\mathcal{U}_i = \text{exp}(\mathfrak{g}_i)$ in the leading order is
\begin{equation}
     \underset{U \sim \mu_H}{\mathbb{E}}\!\left(\mathcal{F}\left[\ket{\psi_0}, G, \mathcal{H}_i\right]\right)
    \;\approx \; \frac{4\,\mathrm{Tr}\!\left[(\Pi_i G \Pi_i)^2\right]}{d_i}.
\end{equation}
\noindent\textit{Proof.---} The adjoint of the generator, $\text{ad}_{U}\left( G\right) = \overline{U}GU$ acts as the phase embedding operator for the protocol. 
We define $\mathcal{M}_{\mu_H}^{(k)}(O) = \underset{U \sim \mu_H}{\mathbb{E}} \left( U^{\otimes k}O\overline{U}^{\otimes k}\right)$ as the $k$-th moment operator of the Haar distribution over the unitary group $\mathcal{U}_i$. 
In terms of these, the average of the Fisher information over $\mathcal{U}_i$ is 
$
        \underset{U \sim \mu_H}{\mathbb{E}}\!\left(\mathcal{F}\left[\ket{\psi_0}, G, \mathcal{H}_i\right]\right) = 4  \bra{\psi_0}\left[ \mathcal{M}_{\mu_H}^{(1)}(G_i^2) - \mathcal{M}_{\mu_H}^{(2)}(G_i^{\otimes 2})\right]\ket{\psi_0} 
$, where $G_i = \Pi_iG\Pi_i$.  
It can be shown using Weingarten calculus \cite{mele2024introduction} that the leading order contribution to this expression is due to the orthogonal projection of $G_i^2$ onto the first order commutant of $\mathcal{U}_i$ such that $\underset{U \sim \mu_H}{\mathbb{E}}\!\left(\mathcal{F}\left[\ket{\psi_0}, G, \mathcal{H}_i\right]\right)
    \;\approx \; 4\,\mathrm{Tr}\!\left[(\Pi_i G \Pi_i)^2\right]/d_i$. \hfill$\square$ \\
See Appendix~\ref{Haar_Ramsey_average_Fisher} for a detailed derivation.

Notably, the metrological performance is a result of both the projection of $G_i^2$ and the dimension of the subspace in which its effective orbit is restricted. 
For the full algebra $\mathfrak{g} = \mathfrak{su}(2^n)$, the QFI is shot-noise limited due to curse of dimensionality \cite{Paolo}. 
This behaviour can be explained using the idea of metrologically sensitive operator orbits introduced above.
For an arbitrary Pauli string, the fraction of operator orbits in which it appears in linear combination with its corresponding metrologically compatible subspace is exponentially suppressed. 
This suppression can be understood as arising from the high cardinality of $\mathcal{C}_{n/2}$, 
which reflects the high degeneracy of $G$ at the middle of the spectrum in the full space. 

However, when the dynamics is restricted to the permutationally invariant subalgebra, $\mathfrak{g} = \mathfrak{su}(n + 1) \subset \mathfrak{su}(2^n)$, the 
average QFI scales as $\mathcal{O}(n^2)$, exhibiting Heisenberg scaling. 
This is well understood using the idea of \textit{symmetrized Pauli strings} \cite{kazi2024universality}, defined as the sum of all permutations of a Pauli string 
\begin{equation}
    B_{\vec{p}} = \mathcal{T}_{S_n}\left( X^{p_X}\otimes Y^{p_Y}\otimes Z^{p_Z}\otimes I^{p_I} \right),
\end{equation}
where $\vec{p} = \left( p_X, p_Y, p_Z, p_I\right)$ represents the contributions of the individual Pauli operators such that $\sum_{\alpha} p_{\alpha} = n$.
These symmetrized Pauli strings now stand in for the operators constituting the permutationally invariant unitary orbits and therefore replace the individual Pauli strings in the definition of equivalence classes. 
In doing so, the cardinality of the individual equivalence classes is reduced substantially (see Appendix~\ref {equi_ramsey} for details).  
This confines the effective orbit of $G$ to a smaller subspace. 
For a generic Pauli string, linear combinations with elements of its metrologically compatible subspace occur with high probability.  Hence, these yield operator orbits that are typically metrologically sensitive. 
As a consequence, it creates probe states that preserve coherence between macroscopically separated $G$ eigenstates, thereby increasing the ensemble's average QFI. 
We also show that this protocol saturates the quantum Cram\'{e}r-Rao bound for multiparameter estimation as a direct calculation shows that the ratio of the diagonal to the off-diaganal elements of the QFI matrix scales with the subsystem number (see Appendix~\ref{sec:multi_haar}). 
These insights provide a causal explanation for several recent results \cite{oszmaniec_random_2016, shi2024quantum, fiderer2018quantum}. 

Twist-untwist protocols can be understood within this formalism. The probe state here is prepared using the composing Hamiltonian, $H_c = \chi S_x^2$ at a time $t \propto 1/\sqrt{n}$, which is then phase embedded using $G = -S_y$ \cite{davis_approaching_2016}. This constrains the dynamics in a permutationally symmetric subspace, while creating metrologically sensitive operator orbits.  
Finally, to understand that metrology schemes do not benefit solely from restricting the Lie algebra using symmetries, we briefly discuss the c-SYK model which can be expressed in the spin basis using a Jordan-Wigner transformation \cite{gu2020notes}.  
The c-SYK model features a highly nonlocal, number-conserving algebra that block-diagonalizes the Hilbert space, thereby reducing the effective size of the associated dynamical Lie algebra.  
However, it is not metrologically useful as a composing Hamiltonian under phase embedding by $S_z = \sum_{i = 1}^{n}Z_i/2$ because they commute with each other. 
In other words, the evolution of $G$ does not create metrologically useful operator orbits as the dynamics is restricted to a single equivalence class, namely $\mathcal{C}_0$. 
For the fiducial state $\ket{0}^{\otimes n}$, this results in a probe state that is incapable of inducing macroscopic variance in the generator $G$, thereby resulting in $\mathcal{F} \approx 0$.  These insights about metrologically sensitive operator orbits can be used to define a novel resilient design of sensors, wherein there are no explicit symmetries and the random unitaries are drawn from $\text{SU}(2^n)$. Such sensors rely on inducing sensitive orbits using postselection as discussed below.

\noindent\textit{Metrology with projected ensembles.---} 
\begin{figure}
    \centering
    \includegraphics[width=\columnwidth]{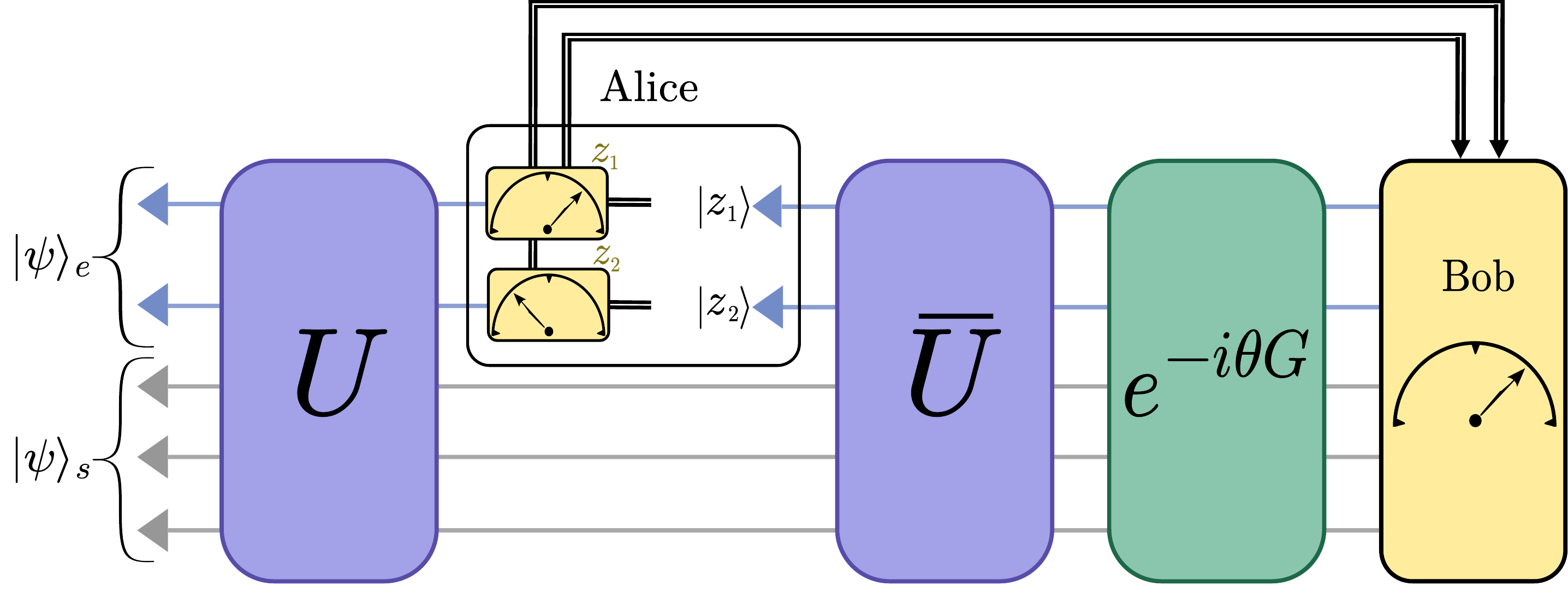}
    \caption{Schematic illustrating the adaptive quantum sensing protocol using projected ensembles. Alice constructs a projected ensemble and communicates the classical measurement outcome to Bob. Bob is able to implement a beyond shot-noise sensor, but only upon use of the measurement outcome.}
\label{fig:projected_ckt}
\end{figure}
Based on the intuition above, we now introduce a new resilient design for quantum sensors, depicted in Fig.~\ref{fig:projected_ckt}. Here we exploit the notion of metrologically compatible subspaces to construct operator orbits that achieve quantum-enhanced performance. 
Consider an $n$-qubit Hilbert space partitioned as $\mathcal{H}=\mathcal{H}_s\otimes\mathcal{H}_e$, where $n_s$ and $n_e$ are the numbers of qubits in the two subsystems, with $n_s \gg n_e$.
A Haar-random unitary $U\in \text{SU}(2^n)$ acts on a fiducial state $\ket{\Psi_0}$ to produce $\ket{\Psi_{se}}$. It is followed by a projective measurement on subsystem $e$ in the computational basis $\{\ket{z_e}\}$ by a party, say Alice. This allows Alice to generate a \textit{projected ensemble} $\mathcal{E}$ of pure states $\ket{\psi_{z_e}} =\braket{z_e|\Psi_{se}}/|\braket{z_e|\Psi_{se}}|$ with probability $p(z_e)=|\braket{z_e|\Psi_{se}}|^2$ on the subsystem $s$ conditioned on the measurement outcome $z_e$~\cite{choi2023preparing, PRXQuantum.4.010311, PRXQuantum.4.030322, bhore, PhysRevLett.134.180403},
\begin{equation}\label{proj_ensemb} 
\mathcal{E}=\bigl\{(p(z_e),\ket{\psi_{z_e}}):z_e\in\{0, 1\}^{\otimes n_e}\bigl\}.
\end{equation}
The measurement outcomes are fed into a state preparation apparatus for the subsystem $e$, after which the time-reversed unitary $\overline{U}$ is applied on the full system. The resulting state acts as the probe state for phase embedding with the generator $S_z$. 
Bob now performs the final two-outcome measurement by first applying $U$, measuring and post-selecting on the outcome $z_e$ provided by Alice, then applying $\overline{U}$ and finally projecting onto the fiducial state $\ket{\Psi_0}$ and its orthogonal component.

It can be shown that this protocol achieves the Heisenberg limit (see Appendix~\ref{sec:projected_ensemble_based_metrology}). 
We note that each step of this protocol is crucial to achieving this quantum-enhanced scaling. 
While it is easy to see the role of measurements in an otherwise insensitive protocol, we note that access to the measurement record also plays a crucial role. 
The measurement record enables access to higher moments of the ensemble, $\rho^{(q)}_\mathcal{E}=\sum_B p(z_e) (\vert\psi_{z_e}\rangle\langle\psi_{z_e}\vert)^{\otimes q}$, in the absence of which Alice sends over an average state for phase embedding which would result in a shot-noise limited protocol. 
As for classical communication of measurement outcomes, it is essential for postselection on Bob’s part; without it, implementing a Fisher-saturating measurement would be difficult. 
Furthermore, we note that the postselection overhead remains low because the number of qubits measured is much smaller than the total system size. 

Here, the local projector corresponding to the measurement at the $i$th site, $\pi_i=(I+Z_i)/2$, naturally fixes $I \in \mathcal{C}_0$ as the reference Pauli string. 
The time reversal operator further scrambles $Z$, generating an operator orbit of the form $(I + \overline{U}Z_iU)/2$. 
The support of this orbit on the equivalence classes is shown in  Fig.~\ref{fig:number_of_excitations}. 
With respect to $I$, $\overline{U}Z_iU$ is a highly scrambled operator with its largest weight in $\mathcal{C}_{n/2}$, which is macroscopically separated from $\mathcal{C}_{0}$ and hence belongs to $\mathfrak{M}_0$, its metrologically compatible subspace. 
As the number of measurement sites $n_e$ increases, the weight of $I$ becomes exponentially suppressed. 
This diminishes its role as a reference Pauli string, and consequently, the scrambled component dominates, approaching a fully Haar–random operator on the entire Hilbert space. 

\begin{figure}
    \centering
    \includegraphics[width=0.85\columnwidth]{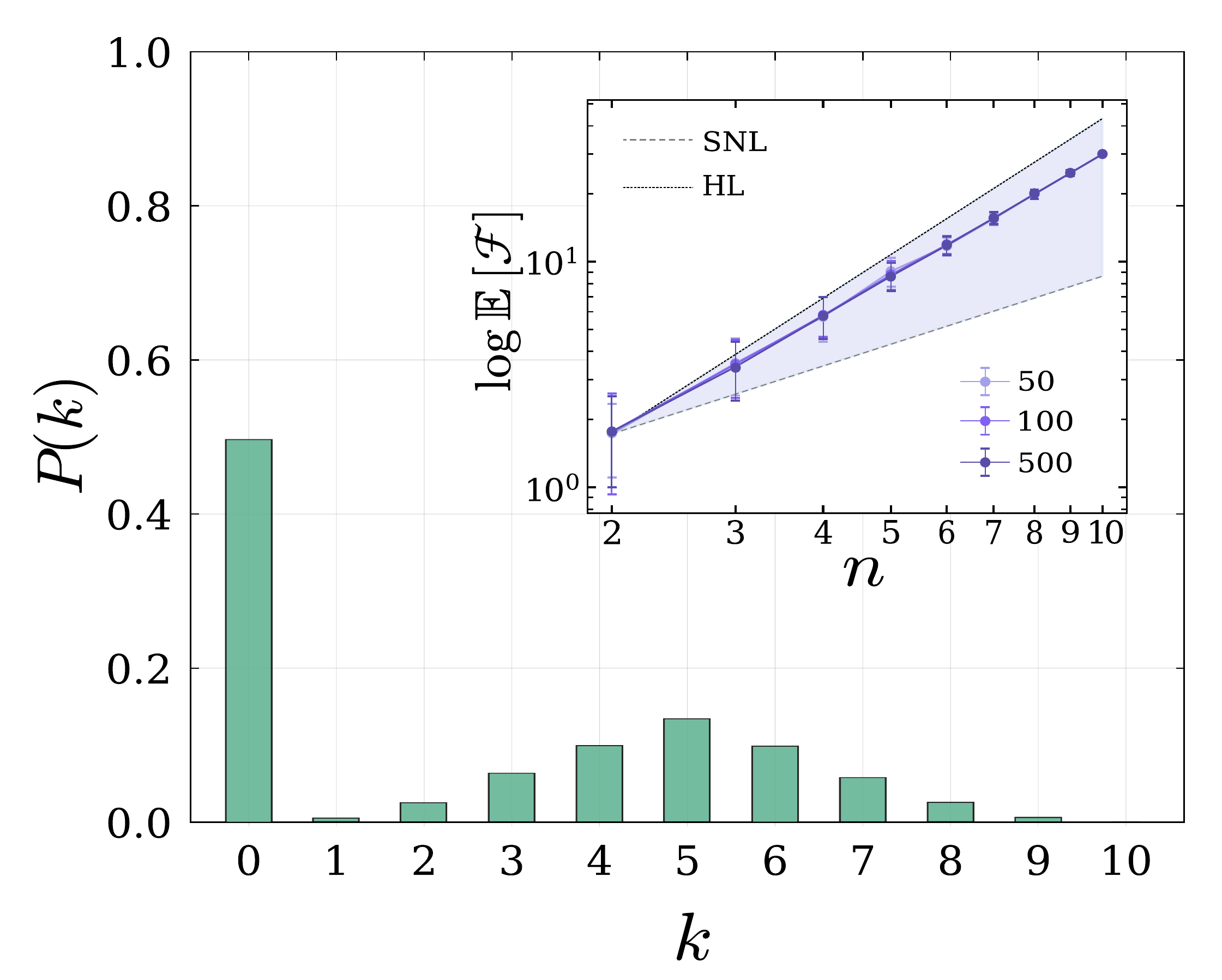}
    \caption{Probability distribution of Pauli strings as a function of their equivalence class index for the protocol using projected ensembles when $n_e = 1$. The inset shows the scaling of Fisher information with the total system size for different sample sizes.}
    \label{fig:number_of_excitations}
\end{figure}

\noindent\textit{Robustness to noise.---}  Besides resilience in design, the efficacy of quantum sensors is informed by their robustness to noise. To this end, we study the effect of noise and particle loss on resilient sensor designs. 
The noise is modeled as acting exclusively on the unitaries in the state preparation and final measurement steps, assuming the encoding dynamics to be effectively decoherence-free due to their short duration. 
Hence, such unitary channels $\mathcal{U}(\cdot) = U(\cdot)\overline{U}$ in these protocols are now replaced with $ \Phi \circ \mathcal{U} (\cdot)$, where $\Phi(\cdot) = \sum_j K_j (\cdot) K_j^\dagger$ is the completely positive trace preserving map modeling noise. 
Specifically, we analyze the impact of depolarizing channels of strength $p$, where $\Phi(\rho) = (1-p)\rho + pI/d$, which is natural in several experimental platforms, including superconducting qubits and photonic systems \cite{PhysRevA.77.012307}.
We denote $\mathcal{F}_{\Phi} \left[ \rho, G, p\right]$ as the QFI for protocols under the impact of such a channel, and $\mathcal{F} \left[ \rho, G\right]$ is the QFI in the absence of noise. 
In the limit of large $d$, the average QFI under the depolarizing channel gets suppressed by a fixed probability with which it retains the original state \cite{sidhu2020geometric}, as can be seen from 
\begin{equation*}
    \underset{U \sim \mu_H}{\mathbb{E}}\left(\mathcal{F}_{\Phi} \left[ \rho, G, p\right]\right) = \frac{(1-p)^2}{\left(1 - p + \frac{2p}{d}\right)} \underset{U \sim \mu_H}{\mathbb{E}}\left(\mathcal{F} \left[ \rho, G\right]\right).
\end{equation*}
Despite this suppression, the characteristic $\mathcal{O}(n^2)$ scaling of the QFI is preserved for both the Haar Ramsey protocol and the protocol using projected ensembles. 

We now evaluate the effect of particle loss at the detector on the metrological performance of our sensors. 
Loss of $k$ particles is modeled as $\rho_{n-k} = \mathrm{Tr}_k\left(\rho_{\text{probe}}\right)$, where $\rho_{\text{probe}}$ is the state of the system before the final measurement onto the fiducial state and its orthogonal component. 
The average QFI for the Haar Ramsey protocol then obeys $\mathbb{E}_{U \sim \mu_H}[\mathcal{F}(\rho_{n-k}, G)] = {(n-k)(n-k+2)}/{3}$. 
For the protocol using projected ensembles, the finite loss rescales the sensitivity as $ \mathcal{F} \sim n^2 e^{-k/2}$ (see Appendix~\ref{sec:particleloss_projected}). 
Therefore, both protocols retain $\mathcal{O}(n^2)$ sensitivity, as long as losses are small enough; however, extensive loss destroys Heisenberg scaling. 

\noindent \textit{Conclusions.---}
We have established a unified framework relating the symmetries of composing and phase-embedding Hamiltonians to the fundamental precision limits of quantum sensors. The interplay between the phase-embedding Hamiltonian and the operator orbit generated by the composing Hamiltonian emerges as the key structure controlling the Fisher information scaling. Random states projected onto symmetry-protected manifolds, such as the permutation-symmetric sector, naturally achieve scaling beyond the shot noise limit.

Building on this framework, we introduced a projected ensemble-based metrology protocol, wherein the effective orbit is reduced through typicality and projection.  
This constitutes an application of our framework, demonstrating that skewing the DLA to enhance overlap with metrologically sensitive operator orbits results in beyond shot noise-limited sensitivity. 
Our results further reveal a close analogy with barren plateaus in variational quantum algorithms \cite{larocca2022diagnosing, larocca2025barren}, identifying an information-theoretic barren plateau in quantum metrology—where precision loss arises not from noise but from unstructured averaging over an exponentially large, symmetry-agnostic unitary space.

From a practical standpoint, these findings carry significant implications for field-deployable quantum sensors. In many real-world applications such as astronomical interferometry, distributed quantum networks, and biological sensing, precisely engineered entangled states are difficult to maintain \cite{zixin, biomed, yukawa2025quantum}. In contrast, partially random or dynamically generated ensembles can emerge naturally through uncontrolled interactions or complex many-body dynamics, making them more experimentally accessible. Examples of such experimental systems that have been driven to chaos include electron and nuclear spins of $^{133}$Cs \cite{jessen}, small Josephson arrays \cite{ergodic_roushan,chavezcarlos}, superconducting devices \cite{google2023measurement} and periodically driven two-component Bose-Einstein condensates \cite{tomkovic}. Our results thus provide a causal and quantitative mechanism to use these existing platforms as quantum-enhanced sensors. In this sense, the framework generalizes quantum metrology beyond laboratory settings, offering a robust, symmetry-informed design principle for scalable, adaptive, and deployable quantum sensing platforms operating under realistic dynamical conditions.

\begin{acknowledgments}
 YI acknowledges the support of the Government of India's DST-INSPIRE and the IITB Asha Navani Foundation. SV acknowledges funding under the Government of India's National Quantum Mission grants numbered DST/QTC/NQM/QC/2024/1 and DST/FFT/NQM/QSM/2024/3. SV thanks P. Zanardi for insight on Haar integration and D. Braun and A. Lakshminarayan for discussions. SV acknowledges the useful discussions at the International Centre for Theoretical Sciences (ICTS) during the programs with codes ICTS/qm1002025/01 and ICTS/qt2025/01. 

\end{acknowledgments}

\newpage

\appendix
\maketitle
\onecolumngrid

\newpage
\section{Time-averaged Quantum Fisher Information for integrable composing Hamiltonians} \label{app: integrable}

In this section, we show that generic integrable Hamiltonians do not exhibit Heisenberg scaling when used as composing Hamiltonians. To do so, we prove that the time average of the phase embedding operator depends only on its invariant components in the highly local algebra generated by the evolution of the integrable Hamiltonian.
\begin{theorem}[Time-averaged QFI and the dynamical Lie algebra]
For a composing Hamiltonian $H_c$ and a phase-embedding operator $G$, let $\mathfrak{g}=\langle\{-\mathrm{i}H_c,\,-\mathrm{i}G\}\rangle_{\text{Lie}}$ denote their dynamical Lie algebra (DLA). 
The finite-time averaged variance of $G$ over a time $T$ in the Heisenberg picture depends only on the component of $G$ or $G^2$ in the centralizer of the DLA, up to corrections that vanish as $1/T$. 
\end{theorem}
\textbf{Proof.}
Let $G(t)=e^{i H_c t} G e^{-i H_c t}$ denote the adjoint evolution of $G$ induced by the composing Hamiltonian. 
Consider the variance of $G(t)$,
\begin{align}
\Delta^{2}_t(G)=\bra{\psi_0}G(t)^2\ket{\psi_0}-\bra{\psi_0}G(t)\ket{\psi_0}^2.
\label{VarG}
\end{align}
The time average of $\Delta^{2}_t(G)$ over $[0,T]$ is
\begin{align}
\overline{\Delta^{2}}_T(G)=\frac{1}{T}\int_0^T \Delta^{2}_t(G)\,\mathrm{d}t.
\end{align}
We assume that the composing Hamiltonian $H_c$ is \emph{integrable}, in the sense that it admits an extensive set of commuting conserved quantities $\{Q_j\}$ \cite{PhysRevB.90.174202}. 
This ensures that the adjoint action $\operatorname{ad}_{H_c}(X)=[H_c, X]$ is diagonalizable with a purely discrete spectrum of Bohr frequencies corresponding to energy differences among the eigenstates of $H_c$. Consequently, the time average over $G(t)$ selects precisely the component of $G$ lying in the kernel of $\operatorname{ad}_{H_c}$,
\begin{align}
\lim_{T\to\infty} \frac{1}{T}\int_0^T G(t)dt = \mathcal{P} (G),
\end{align}
where $\mathcal{P}$ is the projection onto $\ker(\operatorname{ad}_{H_c})$, the centralizer of $H_c$. Decomposing $G$ into eigenoperators of this adjoint action, $G=\sum_\lambda G_\lambda, \operatorname{ad}_{H_c}(G_\lambda)=\lambda G_\lambda,$
 isolates the stationary part $G_0$ (which commutes with $H_c$) from oscillating terms $G_\lambda$ with $\lambda\neq0$.  
Let $\mathcal P$ denote the projection onto the stationary subspace $\ker(\operatorname{ad}_{H_c})$, the centralizer of $H_c$ in $\mathfrak{g}$.  
Then
\begin{align}
G(t)=\mathcal P (G)+\sum_{\lambda\neq0} e^{\mathrm{i}\lambda t} G_\lambda.
\end{align}
Substituting into equation (\ref{VarG}), the stationary term yields a constant variance while the oscillatory cross-terms average to small values,
\begin{align}
\frac{1}{T}\int_0^T e^{\mathrm{i}\lambda t}\,\mathrm{d}t
=\frac{e^{\mathrm{i}\lambda T}-1}{\mathrm{i}\lambda T}.
\end{align}
If $\delta>0$ denotes the smallest nonzero $|\lambda|$ appearing in the decomposition of $G$ or $G^2$, the remainder satisfies $|R_T|\le C/(T\delta)$ for some constant $C$.  
Thus,
\begin{align}
\overline{\Delta^{2}}_T(G)
=\bra{\psi_0}\mathcal P(G)^2\ket{\psi_0}
-\bra{\psi_0}\mathcal P(G)\ket{\psi_0}^2 + R_T.
\end{align}
Setting $A=\bra{\psi_0}\mathcal P(G)^2\ket{\psi_0}, D=\bra{\psi_0}\mathcal P(G)\ket{\psi_0},$
gives
\begin{align}
\overline{\Delta^{2}}_T(G)=A-D^2+R_T,\qquad |R_T|\le \frac{C}{T\delta}.
\end{align}
Finally, since $\mathcal{F}(t)=4\,\Delta^{2}_t(G)$ for pure states,
\begin{align}
[\overline{\mathcal{F}}]_T=4(A-D^2+R_T),
\end{align}
with the same asymptotic scaling.

This implies that 
only this stationary component survives in the long-time average, while noncommuting parts oscillate and decay with a rate set by the adjoint spectral gap $\delta$.

As an illustration, consider the fiducial state  $\ket{\psi_0} = \ket{0}^{\otimes n},$
with generating Hamiltonian   $G = S_z$ and composing Hamiltonian $ H_c = X^{\otimes n} $. The state at time $t$ is
    $\ket{\psi(t)} = \cos(t)\ket{0}^{\otimes n} - i\sin(t)\ket{1}^{\otimes n}.$
Since the evolution is periodic with period $\pi$, the variance oscillates as
    $\Delta^{2}_t(G) = {n^2}\sin^2(2t)/4,$
yielding a time-averaged variance over one period, $\overline{\Delta^{2}}_T(G)
    = \frac{1}{T}\int_0^T \Delta^{2}_t(G)\,dt 
    = \frac{n^2}{8} + {\mathcal O}(1/T).$
The corresponding time-averaged QFI is
    $\overline{\mathcal{F}}_T = 4\,\overline{\Delta^{2}}_T(G)
    = \frac{n^2}{2} + {\mathcal O}(1/T),$
thus retaining the Heisenberg scaling ($\propto n^2$), albeit with a reduced prefactor due to temporal averaging. 

In contrast, consider the same generator and fiducial state, but with a different composing Hamiltonian, $G = \tfrac{1}{2}\sum_{j=1}^n Z_j, 
H_c = \sum_{j=1}^n X_j.$
The DLA generated by $\{-\mathrm{i}H_c, -\mathrm{i}G\}$ factorizes as $\mathfrak g = 
\bigoplus_{j=1}^n 
\langle\{-\mathrm{i}X_j, -\mathrm{i}Z_j\}\rangle_{\text{Lie}}
= \bigoplus_{j=1}^n \mathfrak{su}(2)_j,$
a direct sum of local algebras.  
Each qubit evolves independently under its own adjoint orbit,
\begin{align}
G(t) = e^{\mathrm{i}tH_c}G\,e^{-\mathrm{i}tH_c}
= \tfrac{1}{2}\sum_{j=1}^n
\big(Z_j\cos(2t) + Y_j\sin(2t)\big),
\end{align}
with local frequencies $\lambda_j = \pm 2$.
The projection $\mathcal P$ onto the kernel of $\operatorname{ad}_{H_c}$ 
removes all oscillating components, yielding
$\mathcal P (G) = 0$ but 
$\mathcal P(G^2) = \tfrac{1}{4}\sum_j I_j$. For $\ket{\psi_0}=\ket{0}^{\otimes n}$,
$\langle G(t)\rangle = 0, 
\langle G(t)^2\rangle = \tfrac{n}{4},$
so that $\overline{\Delta^{2}}_T(G)
= \tfrac{n}{4} + \mathcal O(1/T),
[\overline{\mathcal{F}}]_T = \mathcal O(n).$

\section{Haar Ramsey protocol}
In this section, we prove Theorem 1, and present examples for it. 
We also explain these examples using the concept of equivalence classes. 
Further, we show that the Haar Ramsey protocol saturates the QFI for the case of two-parameter estimation.

\subsection{Average Fisher Information}
\label{Haar_Ramsey_average_Fisher}

In this section, we derive Theorem~1.
Consider the unitary group 
$\mathcal{U}_i = \exp(\mathfrak{g}_i),$
equipped with the Haar measure $\mu_H$.  
Under the adjoint action $G \mapsto \overline{U} G U$, the Haar--averaged QFI reads
\begin{align}
\label{eq:Haar_avg_F}
    \underset{U \sim \mu_H}{\mathbb{E}}\!\left(\mathcal{F}\!\left[\ket{\psi_0},G,\mathcal{H}_i\right]\right)
    &= 4\,\bra{\psi_0}\!\left[
        \mathcal{M}^{(1)}_{\mu_H}(G^2)
        -
        \mathcal{M}^{(2)}_{\mu_H}(G^{\otimes 2})
    \right]\!\ket{\psi_0},
\end{align}
where the $k$th Haar moment is
$\mathcal{M}^{(k)}_{\mu_H}(O)
=
\underset{U\sim\mu_H}{\mathbb{E}}\!\left(U^{\otimes k} O\, \overline{U}^{\otimes k}\right).$ By Schur’s lemma \cite{hall2013lie}, the first moment projects onto the identity component of the commutant:
\begin{align}
\mathcal{M}^{(1)}_{\mu_H}(O)
=
\frac{{\rm Tr}(\Pi_i O \Pi_i)}{d_i} .
\end{align}

For the second moment, the commutant of the adjoint action on 
$\mathcal{H}_i^{\otimes 2}$ is spanned by 
$\{\Pi_i\otimes\Pi_i,\mathbb{F}_i\}$, where $\mathbb{F}_i$ denotes the swap operator.  
Hence
\begin{equation}
\label{eq:2ndmoment}
    \mathcal{M}^{(2)}_{\mu_H}(O)
    =
    c_{\Pi_i,O}\,(\Pi_i\!\otimes\!\Pi_i)
    +
    c_{\mathbb{F}_i,O}\,\mathbb{F}_i ,
\end{equation}
with Weingarten coefficients
\begin{align}
c_{\Pi_i,O}
=
\frac{{\rm Tr}(\Pi_i O) - d_i^{-1}{\rm Tr}(\mathbb{F}_iO)}
{d_i^{2}-1},
\qquad
c_{\mathbb{F}_i,O}
=
\frac{{\rm Tr}(\mathbb{F}_i O) - d_i^{-1}{\rm Tr}(\Pi_i O)}
{d_i^{2}-1}.
\end{align}
Note that $G$ is within a given irreducible block. Therefore,  substituting \eqref{eq:2ndmoment} into \eqref{eq:Haar_avg_F}, we obtain the leading-order expression
\begin{align}
\underset{U \sim \mu_H}{\mathbb{E}}\!\left[\mathcal{F}(\ket{\psi_0},G, \mathcal{H}_i)\right]
\approx
\frac{4\,{\rm Tr}\!\left[(\Pi_i G\Pi_i)^2\right]}{d_i},
\end{align}
which completes the proof of Theorem~1. We now illustrate the general Haar–averaged expression with several physically relevant ensembles:

\begin{enumerate}

\item[(a)] \textbf{Full unitary group.}  
For $\mathcal{E}=SU(d)$ acting irreducibly on $\mathcal{H}$ (so $\Pi_i=I$ and $d_i=d$), and for traceless generators $G$, Schur’s lemma gives
\begin{align}
\mathcal{M}^{(1)}_{\mu_H}(G^2)=\frac{{\rm Tr}(G^2)}{d}\,I,
\qquad
\mathcal{M}^{(2)}_{\mu_H}(G^{\otimes 2})=\frac{{\rm Tr}(G^2)}{d(d^2-1)}\left(I-\frac{\mathbb{F}}{d}\right).
\end{align}
Substitution into the Haar–averaged QFI yields
\begin{align}
\underset{U \sim \mu_H}
{\mathbb{E}}\!\left(\mathcal{F}[\ket{\psi_0},G]\right)
\approx
\frac{4\,{\rm Tr}(G^2)}{d}.
\end{align}
For collective spin generators with ${\rm Tr}(G^2)=n2^n$ and $d=2^n$, one obtains only linear scaling in $n$.

\item[(b)] \textbf{Symmetric (permutationally invariant) subspace.}  
Let $\Pi_i$ be the projector onto the symmetric irrep of $n$ qubits, with dimension $d_i=2S+1=n+1$.  
The generator restricted to this sector is $G_i=\Pi_i G \Pi_i$.  
For $G=S_z$,
\begin{align}
{\rm Tr}(G_i^2)
=
{\rm Tr}[(\Pi_i G\Pi_i)^2]
=
\frac{S(S+1)(2S+1)}{3}
\sim\frac{2S^3}{3}.
\end{align}
Applying the leading-order Haar–average derived previously,
\begin{align}
\underset{{U\in\mathsf{U}(d_i)}}{\mathbb{E}}\!\left(\mathcal{F}[\ket{\psi_0},G, \mathcal{H}_i]\right)
\approx
\frac{4\,{\rm Tr}(G_i^2)}{d_i}
=
\frac{8S^3}{3(2S+1)}
\sim\mathcal{O}(S^2)\sim\mathcal{O}(n^2).
\end{align}

\item[(c)] \textbf{Orthogonal ensemble on the symmetric subspace.}  
Let $\mathcal{E}=O(d_i)$ act on the symmetric irreducible representation $\mathcal{H}_i$.  
Since $\overline{O}=O^T$, the second commutant of $O(d_i)$ on $\mathcal{H}_i^{\otimes 2}$ is spanned by  
\begin{align}
\{\Pi_i\otimes\Pi_i,\ \mathbb{F}_i,\ |{\Omega_i}\rangle\langle{\Omega_i}| \},
\qquad
\ket{\Omega_i} = (\Pi_i\otimes\Pi_i)\sum_j \ket{jj},
\end{align}
yielding the decomposition
\begin{align}
\mathcal{M}^{(2)}_{O(d_i)}(G_i^{\otimes 2})
=
c_{\Pi_i}\,(\Pi_i\!\otimes\!\Pi_i)
+
c_{\mathbb{F}_i}\,\mathbb{F}_i
+
c_{\Omega_i}\,|{\Omega_i}\rangle\langle {\Omega_i}|,
\end{align}
with coefficients determined by  
${\rm tr}(G_i)$, ${\rm tr}(G_i^2)$, and ${\rm tr}(G_i^T G_i)$.
For a collective generator $G=S_x$, one has
\begin{align}
{\rm tr}(G_i^2)={\rm tr}(G_i^T G_i)=\frac{S(S+1)(2S+1)}{3}\sim\frac{2S^3}{3}.
\end{align}

Using the general formula
\begin{align}
\mathbb{E}_{O\sim\mu_H}\!\left[\mathcal{F}\right]
=
4\!\left(
\frac{{\rm tr}(G_i^2)}{d_i}
-
\bra{\psi_0}
\mathcal{M}^{(2)}_{O(d_i)}(G_i^{\otimes 2})
\ket{\psi_0}
\right),
\end{align}
and the above commutant decomposition, one obtains
\begin{align}
\mathbb{E}_{O\sim\mu_H}\!\left[\mathcal{F}(\ket{\psi_0},G)\right]
=
\frac{4\,{\rm tr}(G_i^2)}{d_i}
-
\frac{4\!\left[{\rm tr}(G_i^2)+{\rm tr}(G_i^T G_i)+{\rm tr}(G_i)^2\right]}{(d_i+2)d_i}.
\end{align}
For traceless $G_i$ and large $d_i$,
\begin{align}
\mathbb{E}_{O\sim\mu_H}\!\left[\mathcal{F}\right]
\approx
4\!\left(
\frac{{\rm tr}(G_i^2)}{d_i}
-
\frac{{\rm tr}(G_i^T G_i)+{\rm tr}(G_i^2)}{d_i^2}
\right)
\approx
\frac{8S^3}{3d_i}.
\end{align}
For $d_i=2S+1$, this again yields  
\begin{align}
\mathbb{E}_{O\sim\mu_H}\!\left[\mathcal{F}\right]\sim\mathcal{O}(S^2).
\end{align}

\end{enumerate}
\hfill$\square$

\subsection{Equivalence classes} \label{equi_ramsey}
We now provide explanations supporting the definition in Eq.~(\ref{eq:eq_class}). 
For the generator $G = S_z = \tfrac12\sum_{i=1}^n Z_i$ and a Pauli string $P=\bigotimes_{i=1}^n P_i$, one can define an ``anticommutation count", $N(P,S_z) = \#\{i : P_i \in \{X,Y\}\}$. 
Two Pauli strings $P$ and $Q$ are said to be equivalent to one another if $N(P, S_z) = N(Q, S_z)$. 
Therefore, the equivalence class $\mathcal{C}_k$ consists of all strings with exactly $k$ locally anticommuting sites. 
Its size is $|\mathcal C_k|=\binom{n}{k}2^n$, giving normalized weight $f_k=\binom{n}{k}/2^n$ among all $4^n$ Pauli strings. 
Thus, $N(P, S_z) \sim \mathrm{Binomial}(n,\tfrac12)$ and $\Pr(|N(P, S_z) -n/2|\ge\varepsilon n)\le 2e^{-2\varepsilon^2 n}$, implying exponential concentration near $k = n/2$ \cite{chernoff1952measure}. 
For a reference $P\in \mathcal C_p$, the metrologically compatible subspace $\mathfrak M_p=\mathrm{span}\bigl(\cup_{\lvert k-p\rvert=\mathcal{O}(n)}\mathcal C_k\bigr)$ contains operators whose commutation pattern differs from $P$ on a macroscopic number of sites. Nonetheless, a random Pauli lands in such classes with exponentially small probability. Typical operators are concentrated at $k\approx n/2$, making any linear combination between $P$ and $\mathfrak M_p$ exponentially suppressed.

Therefore the orbit of $S_z$ in $\mathfrak{su}(2^n)$ contains almost no components that are metrologically sensitive, and the Haar–averaged QFI remains shot-noise limited, scaling as $\mathcal O(n)$, reflecting a curse of dimensionality. 

\subsubsection*{B. Permutationally Invariant Algebra}

A qualitatively different behavior emerges upon restricting attention to the permutationally invariant subalgebra $\mathfrak g = \mathfrak{su}(n+1) \subset \mathfrak{su}(2^n),$
generated by \emph{symmetrized Pauli strings}
    $B_{\vec p}
    =
    \mathcal T_{S_n}\!\left(
        X^{p_X} \otimes Y^{p_Y}
        \otimes Z^{p_Z} \otimes I^{p_I}
    \right),
     \sum_{\alpha} p_\alpha = n,$ 
where $\mathcal T_{S_n}$ denotes averaging over all permutations. The number of such operators equals the number of weak compositions of $n$ into four parts \cite{kazi2024universality, sauvage2024classicalshadowssymmetries},
\begin{equation}
    \#\{ B_{\vec p} \}
    = \binom{n+3}{3}
    = \mathcal O(n^3),
\end{equation}
which is polynomial rather than exponential in $n$.

For a fixed anticommutation count $k=p_X+p_Y$, the number of symmetrized operators is
\begin{equation}
    \bigl|\{ B_{\vec p} : p_X + p_Y = k \}\bigr|
    = (k+1)(n-k+1),
\end{equation}
so the normalized weight is
\begin{equation}
    \tilde f_k
    = \frac{(k+1)(n-k+1)}{\binom{n+3}{3}}.
\end{equation}
Unlike the binomial distribution $f_k$ in the full algebra, the distribution $\tilde f_k$ is \emph{not} exponentially concentrated around $k=n/2$. The probability of sampling an operator with $|k-n/2|=\mathcal{O}(n)$ is only polynomially suppressed (typically of order $1/n$), not $e^{-\mathcal{O}(n)}$.

Hence, for a typical symmetrized Pauli string, the overlap with its metrologically compatible subspace $\mathfrak M_p$ is \emph{typical} rather than exponentially unlikely. The resulting operator orbits now possess macroscopic support across widely separated $S_z$ eigenvalues, allowing the construction of probe states with 
$
    \mathrm{var}(S_z) = \mathcal{O}(n^2),
$
and therefore
    $\mathcal F = \mathcal O(n^2),$ achieving Heisenberg scaling for the average QFI.

\subsection{Two-parameter estimation}\label{sec:multi_haar}
Here we show that Haar-Ramsey protocol saturates the quantum Cram\'{e}r-Rao bound for multiparameter estimation. In this section, we generalize the single-parameter estimation protocol to two-parameter estimation, relevant to applications such as vector magnetometry, where the parameter-encoding Hamiltonian takes the form
\begin{align}
G =\theta_1 G_1 + \theta_2 G_2.
\end{align}
Here, $\theta_1$ and $\theta_2$ denote the parameters to be estimated and $G_1$, $G_2$ represent the associated Hermitian generators. 
In a local estimation scenario, where $\theta_{1,2} \ll 1$, the unitary evolution $e^{iG}$ can be approximated to second order using the Baker-Campbell-Hausdorff expansion. Specifically,
\begin{subequations}
\begin{align}
e^{i(\theta_1 G_1 + \theta_2 G_2)} &\approx I + i(\theta_1 G_1 + \theta_2 G_2) - \frac{1}{2}(\theta_1^2 G_1^2 + \theta_2^2 G_2^2 + \theta_1\theta_2\{G_1, G_2\}), \\
e^{i\theta_1 G_1} e^{i\theta_2 G_2} &\approx I + i(\theta_1 G_1 + \theta_2 G_2) - \frac{1}{2}(\theta_1^2 G_1^2 + \theta_2^2 G_2^2) - \theta_1\theta_2 G_1 G_2, \\
e^{\theta_1\theta_2[G_1, G_2]/2} &\approx I + \frac{\theta_1\theta_2}{2}[G_1, G_2],
\end{align}
\end{subequations}
which implies, up to $\mathcal{O}(\theta^2)$,
\begin{align}
e^{i\theta_1 G_1} e^{i\theta_2 G_2} e^{\theta_1\theta_2[G_1, G_2]/2} = e^{i(\theta_1 G_1 + \theta_2 G_2)}.
\end{align}
The phase-encoded state is then expressed as
\begin{align}
|\psi\rangle = \overline{U}  e^{i\theta_1 G_1} e^{i\theta_2 G_2} e^{\frac{\theta_1\theta_2}{2}[G_1, G_2]} U \ket{0}, \label{eq:phase_embedded_state}
\end{align}
where $U$ is chosen from an ensemble ${\cal E}$. 
In practical scenarios, we consider $G_1 = \sum_i X^{(i)}$, $G_2 = \sum_i Y^{(i)}$ such that $[G_1, G_2] = 2iG_3.$

To quantify estimation precision, we compute the quantum Fisher information matrix (QFIM), given by
\begin{align}
[\mathcal{F}(\boldsymbol{\theta})]_{jk} = 4\mathrm{Re} \left[ \langle \partial_j \psi | \partial_k \psi \rangle - \langle \partial_j \psi | \psi \rangle \langle \psi | \partial_k \psi \rangle \right].
\end{align}

Substituting the values, it turns out that the dominant contribution in the  diagonal terms of QFIM is given by
\begin{align}
\mathop{\mathbb{E}}\limits_{{U \sim \mu_H}}[\mathcal{F}]_{11} =\mathop{\mathbb{E}}\limits_{{U \sim \mu_H}} [\mathcal{F}]_{22} = \frac{{\rm Tr} G^2}{d},
\end{align}
where $G$ may be chosen to be either $G_1$ or $G_2$ and the off-diagonal terms are vanishingly small. 

As in the single-parameter estimation scenario, if the unitaries are sampled from $\text{SU}(2^n)$, the Haar-averaged Fisher information scales only linearly with the system size. 
However, if the sampling is restricted to a permutationally symmetric subspace of $\text{SU}(2^n)$, the Haar-averaged Fisher information grows quadratically in the system size. These expressions yield the complete Haar-averaged QFIM for the two-parameter estimation problem.

\section{Metrology with Projected Ensembles}
\label{sec:projected_ensemble_based_metrology}

Here, we prove that the projected ensemble protocol produces Heisenberg-limited sensors. Consider a bipartite system of $n$ qubits prepared in a globally random pure state
$\ket{\Psi_{se}} = U \ket{0}^{\otimes n}, U = e^{-i H_c t} \in \text{SU}(2^n)$. Here, $H_c$ is an entangling Hamiltonian drawn from a suitable ensemble.  
A projective measurement on subsystem $e$ in the computational basis produces conditional pure states on $s$,
\begin{align}
\ket{\psi_{z_e}} &= \frac{(\bra{z_e}\otimes I_s)\ket{\Psi_{se}}}{\sqrt{p(z_e)}}, \qquad 
p(z_e) = \|(\bra{z_e}\otimes I_s)\ket{\Psi_{se}}\|^2,
\end{align}
defining the \emph{projected ensemble}
$\mathcal{E} = \{(p(z_e), \ket{\psi_{z_e}})\}$.
Phase estimation is performed via the collective generator
\begin{align}
S_z = \tfrac{1}{2}\sum_{i=1}^{n_A} Z_i.
\end{align}

The action of $S_z$ on any operator $X$ defines the adjoint map $\mathrm{ad}_{S_z}(X) = [S_z, X]$.  

\paragraph{Restoring metrological support via projection and untwirling}

A modified protocol circumvents the QFI suppression in $\text{SU}(2^n)$ by introducing an additional step that aligns the orbit within the metrologically sensitive subspace.  
Specifically, consider the operator
$G' \;=\; U\,\pi\,\overline{U}\,G\,U\,\pi\,\overline{U},$
where $\pi = I + A$ is a local or few-body projector, $A$ is a Pauli string, and $\tilde{\pi} = U\pi\overline{U}$.  
This operation effectively selects a reference Pauli for the orbit and confines the resulting support to metrologically compatible subspace.  
Writing explicitly,
$G' = \tilde{\pi} G \tilde{\pi} 
      = \sum_i \lambda_i\,\tilde{\pi}\ket{\lambda_i}\bra{\lambda_i}\tilde{\pi}
      = \sum_i \lambda_i (I+\tilde{A})\ket{\lambda_i}\bra{\lambda_i}(I+\tilde{A})$.    

In physical terms, the projection $\pi$ fixes $I$ as a \emph{reference Pauli} such that $G'$ is now a metrologically sensitive operator orbit.  
Consequently, the resulting Fisher information exhibits superextensive scaling, $
\mathcal{F}(\psi, S_z) \sim \mathcal{O}(n^2),
$ recovering Heisenberg-limited metrological enhancement despite initial randomization.

\subsection{Effect of particle loss} \label{sec:particleloss_projected}
We now consider particle loss for the sensors based on projected ensembles. We consider an $n$-qubit symmetric system expressed in the Dicke basis. 
Let $\ket{0_n}$ denote the fully unexcited state and $\ket{q_n}$ the Dicke state with $q = n/2$ excitations. 
The initial state is
\begin{equation}
    \ket{\psi} = \alpha \ket{0_n} + \beta \ket{q_n}, 
    \qquad
    |\alpha|^2 + |\beta|^2 = 1.
\end{equation}
We evaluate the quantum Fisher information (QFI) for phase shifts generated by $S_z$, considering two cases: 
(i) loss of $k$ particles before phase encoding, and 
(ii) loss after encoding.

\subparagraph{(i) Particle loss before phase embedding.---}
After tracing out $k$ particles, the remaining subsystem of $r = n - k$ qubits is described by
\begin{equation}
\ket{q_n} 
\;\longrightarrow\;
\sum_{j=0}^{r} \sqrt{p_j}\, \ket{j_r}\ket{q-j}_k,
\qquad
p_j = \frac{\binom{r}{j}\binom{k}{q-j}}{\binom{n}{q}}.
\end{equation}
The reduced density matrix of the remaining $r$ qubits reads
\begin{align}
\rho_r =&\;
a\, \ket{0_r}\bra{0_r}
+ b\, \ket{q_r}\bra{q_r}
+ \sum_{j\neq 0, q} |\beta|^2 p_j \ket{j_r}\bra{j_r}+ c\, \ket{0_r}\bra{q_r} + c^* \ket{q_r}\bra{0_r},
\end{align}
where
\begin{align}
    a = |\alpha|^2 + |\beta|^2 p_0^{(\mathrm{from}\, q)}, 
    b = |\beta|^2 p_q^{(\mathrm{remain})}, c = \alpha \beta^* s,
\end{align}
and
\begin{align}
    p_q^{(\mathrm{remain})} &= \frac{\binom{r}{q}}{\binom{n}{q}}, \qquad
    p_0^{(\mathrm{from}\, q)} = \frac{\binom{k}{q}}{\binom{n}{q}}, \qquad
    s = \sqrt{p_q^{(\mathrm{remain})}}.
\end{align}
If $q > r$, i.e., more than half of the particles are lost, the coherence vanishes ($s=0$).

Each Dicke state $\ket{j_r}$ is an eigenstate of $S_z$ with eigenvalue $h_j = j - r/2$, so the relevant eigenvalue difference is $\Delta h = q$. 
Restricting to the subspace $\mathrm{span}\{\ket{0_r}, \ket{q_r}\}$, the QFI becomes
\begin{equation}
    \mathcal{F}[\rho_r; S_z]
    = \frac{4(\Delta h)^2 |c|^2}{a + b}
    = \frac{4 q^2 |\alpha|^2 |\beta|^2\, p_q^{(\mathrm{remain})}}
    {|\alpha|^2 + |\beta|^2\!\left[p_0^{(\mathrm{from}\, q)} + p_q^{(\mathrm{remain})}\right]}.
\end{equation}
For $k=0$, the state is pure and $\mathcal{F} = 4q^2|\alpha|^2|\beta|^2$, whereas for $k > n/2$, $\mathcal{F} = 0$.

\subparagraph{(ii) Particle loss after phase embedding.---}
After the phase encoding,
\begin{equation}
\ket{\Psi(\varphi)} 
= \alpha e^{i\varphi n/2}\ket{0_n} + \beta \ket{q_n},
\end{equation}
up to a global phase. 
Tracing out $k$ particles yields
\begin{equation}
\rho_r(\varphi) =
\begin{pmatrix}
a & c\, e^{i\theta}\\
c^* e^{-i\theta} & b
\end{pmatrix}_{\{\ket{0_r},\,\ket{q_r}\}}
\oplus (\text{other diagonal terms}),
\qquad
\theta = \frac{n}{2}\varphi.
\end{equation}
For this two-level subsystem, the QFI with respect to $\theta$ is $\mathcal{F}[\rho_r(\theta); \theta] = 4|c|^2/(a+b)$. 
Using $(\mathrm{d}\theta/\mathrm{d}\varphi)^2 = (n/2)^2$, we obtain
\begin{equation}
\mathcal{F}[\rho_r(\varphi); \varphi]
    = \Big(\frac{n}{2}\Big)^2 \frac{4|c|^2}{a+b}
    = \frac{n^2 |\alpha|^2 |\beta|^2\, p_q^{(\mathrm{remain})}}
    {|\alpha|^2 + |\beta|^2\!\left[p_0^{(\mathrm{from}\, q)} + p_q^{(\mathrm{remain})}\right]}.
\end{equation}

The QFI is therefore identical whether particle loss occurs before or after phase encoding. 
Hence, for this two-component Dicke superposition, particle loss and phase embedding commute with respect to metrological sensitivity. 
For $k=0$, $\mathcal{F} = n^2 |\alpha|^2 |\beta|^2 = 4q^2|\alpha|^2|\beta|^2$, while for $k > n/2$, coherence is lost and $\mathcal{F} = 0$. 
The general expression can be written compactly as
\begin{equation}
\mathcal{F}
= \frac{n^2 |\alpha|^2 |\beta|^2\, p_q^{(\mathrm{remain})}}
{|\alpha|^2 + |\beta|^2\!\left[p_0^{(\mathrm{from}\, q)} + p_q^{(\mathrm{remain})}\right]},
\quad
p_j = \frac{\binom{r}{j}\binom{k}{q-j}}{\binom{n}{q}},\;
r = n - k,\;
q = \frac{n}{2}.
\end{equation}

\subparagraph{Limiting behavior.---}

\emph{No loss ($k=0$).---}
For $r=n$, one has $p_q^{(\mathrm{remain})}=1$ and $p_0^{(\mathrm{from}\,q)}=0$, giving
$\mathcal{F} = n^2 |\alpha|^2 |\beta|^2 = 4 q^2 |\alpha|^2 |\beta|^2$, corresponding to the Heisenberg limit.
For a balanced superposition ($|\alpha|^2=|\beta|^2=1/2$), $\mathcal{F} = n^2/4$.

\emph{Complete loss or $k>n/2$.---}
If more than half the qubits are lost ($r<q$), the coherence between $\ket{0_r}$ and $\ket{q_r}$ vanishes, $p_q^{(\mathrm{remain})}=0$, and consequently $\mathcal{F}=0$.

\emph{Small fixed loss ($k\ll n$).---}
Using 
$p_q^{(\mathrm{remain})}=\frac{\binom{n-k}{q}}{\binom{n}{q}}
=\prod_{i=0}^{q-1}\frac{n-k-i}{n-i}\approx(1-\frac{k}{n})^q$,
and setting $q=n/2$, one finds
$p_q^{(\mathrm{remain})}\approx(1-\frac{k}{n})^{n/2}\!\to e^{-k/2}$ for large $n$. 
Hence for small $k$, the QFI retains Heisenberg scaling with a constant suppression factor,
\begin{equation}
\mathcal{F} \approx n^2 |\alpha|^2 |\beta|^2 e^{-k/2}.
\end{equation}
For the balanced case, $\mathcal{F} \simeq \frac{n^2}{4}(1 - k/4 + \cdots)$.

\emph{Fractional loss ($k = \gamma n$).---}
For a finite loss fraction $0<\gamma<1$,
\begin{align}
p_q^{(\mathrm{remain})}\approx (1-\gamma)^{n/2}
= \exp\!\left[\tfrac{n}{2}\ln(1-\gamma)\right],
\end{align}
which decays exponentially with $n$. 
Thus $\mathcal{F}$ is exponentially suppressed, and Heisenberg scaling is destroyed when a finite fraction of qubits is lost.

The QFI scales as $\mathcal{F} \propto n^2$ for $k/n\to 0$, but vanishes exponentially for any fixed loss fraction $\gamma>0$. 
Hence, finite particle loss only rescales the sensitivity, while extensive loss irreversibly destroys the metrological advantage.

\end{document}